\documentclass[nohyper,12pt,letterpaper]{JHEP}
\usepackage{graphicx}

\newfont{\frak}{eufm10 scaled 1200}

\newfont{\Bbb}{msbm10 scaled 1200}     
\newcommand{\mathbb}[1]{\mbox{\Bbb #1}}
\DeclareSymbolFont{AMSa}{U}{msa}{m}{n}
\DeclareSymbolFont{AMSb}{U}{msb}{m}{n}
\let\Box\relax
\DeclareMathSymbol{\Box}{\mathord}{AMSa}{"03}

\def \eqn#1#2{\begin{equation}#2\label{#1}\end{equation}}

\def \p{\pi}

\title{The Energy Density of "Wound" Fields in a Toroidal Universe}

\author{ W. Fischler, S. Paban and M. \v{Z}ani\'{c} \\
      Department of Physics\\
      University of Texas, Austin, TX 78712\\

{\tt fischler, paban, zanic @zippy.ph.utexas.edu}}

\abstract{The observational limits on the present energy density
of the Universe allow for a component that redshifts like $1/a^2$
and can contribute significantly to the total. We show that a possible
origin for such a contribution is that the universe has a toroidal
topology with "wound" scalar fields around its cycles.}


\preprint{\hepth{} \\ UTTG-07-04}
\begin{document}


\section{Introduction }

In this note we will describe a rather tight scenario which leads to a
contribution to the present energy density of the universe, at the level
of roughly one percent and redshifts like $1/ a^2$, in addition to
  the cosmological constant and matter.

The script has several actors whose parts we describe next. The
story starts by assuming a flat universe with the spatial topology of a
torus, more specifically, this torus is a cube with opposite sides
identified.

We assume that the universe undergoes a period of inflation which
has the minimum amount of e-foldings necessary to solve the horizon and
flatness problem. Any period of inflation beyond that number of e-foldings
amount would inevitably dilute the $1/a^2$ contribution well below one
percent of the present energy density. The physical size of the
torus at the onset of inflation will be assumed to be such that the size
of this torus today reaches the lower bound obtained by the search for
"circles in the sky"  \cite{Cornish:2003db}. There are other bounds
on the topology of the universe from the CMB fluctuations \cite{deOliveira-Costa:1997fc}
as well as from galactic methods \cite{Uzan:1998hk,Levin:2001fg}.

The matter content has, in addition to the inflaton and conventional low
energy states, at least three massless angle-valued fields, ${\theta}_i$,
similar to axions.

We will show in what follows, that with these actors and some additional
assumptions which will be introduced further in the text,  we will be
lead to the existence of a component in the energy density that scales
like $1/a^2$. Such a contribution to the Friedmann equation mimicks a
negative spatial curvature term but unlike the curvature term, it does
not modify the relation between angular and the radial distance from what
it is in flat space \cite{Caldwell:2004vi}. One crucial assumption 
will be the existence of a
non-trivial topological configuration for the angles, ${\theta}_i$.

One might wonder why anybody would entertain such extreme assumptions.
A partial answer is that the present data, as best we can tell, does
not exclude a contribution of a percent or so to the present energy
density that could arise from such an extravagant construction and
therefore warrants some exploration.

We will begin by clearly stating the various assumptions and then describe
how they can lead to a contribution of the order of a percent to the
total energy density of the present universe.

\section{The Spatial Torus and the Topology of Fields}

We will present in this section the setup that leads to a
one percent contribution to the present energy density of the universe
which exhibits the $1/a^2$ behaviour. In what follows, the
universe is assumed to be spatially flat with the topology of a
torus.

For simplicity, we will take the torus to be a
cube with opposite sides identified. The coordinate distance between
opposite sides of the cube is $L$. We will denote the three cycles of the
torus by
$x_i = \frac{ L}{ 2 \p} {\chi}_i$, where the ${\chi}_i$ are angles.

The next essential ingredient involves the existence
of three minimally coupled scalar fields, ${\varphi}_i$ where
$i=1,2,3$ which span in field space, a cube with opposite sides
identified. This cube\footnote{If one
allows for different values for the "decay constants", $f_i$, then as 
will be seen a little further,
  in
order to respect the limits on the isotropy of the microwave
background, the sizes, $L_i$, of the cycles of the spatial torus must
satisfy $L_1/ w_1f_1 = L_2/ w_2f_2 = L_3/ w_3f_3$} in field space has 
linear size $f$, in our
script $f$ (like $L$) is a parameter available for tuning.

\eqn{thetai}{{\varphi}_i = \frac{f}{2 \p}{\theta}_i}
where the ${\theta}_i$ are angle valued fields.

These scalar fields are assumed to couple to matter through their
derivatives\footnote{The assumption that the ${\varphi}_i$ are minimally
coupled scalars ensures that the gravitational coupling of these
scalars to matter is also through derivatives.} and that no potential is
generated for these fields, i.e. the Lagrangian and the quantum dynamics
that it generates, is invariant under the following transformations:

\eqn{thetatransform}{{\theta}_i \rightarrow {\theta}_i + c_i}
  where $c_i$ are (space-time independent) constants. These scalars can be
thought of as massless axions. In order to protect the isotropy of 
the microwave
background as will become clear next,  we impose an exact $Z_3$ symmetry,
  which acts on the three ${\varphi}_i$.

The equations of motion for the scalar fields in a
Friedmann-Robertson-Walker (FRW) flat universe are:

\eqn{FRWscalar}{\ddot{\theta}_i + 3{\dot{a}\over{a}} \dot{\theta}_i -
{{\nabla}^2 \theta_i\over a^2} = 0}
where the FRW metric we use is:\eqn{frw}{ds^2 = - dt^2 + a^2(t) 
\delta_{ij} dx^i dx^j}

Next, we choose as solution of equations (\ref{FRWscalar}), a
topologically non-trivial configuration for the scalar fields:

\eqn{topol}{{\theta}_i = w {\chi}_i}
where $w$ is the same winding number for each angular field around its
respective cycle.

This very special choice of a background solution for
the ${\theta}_i$, will ensure that we comply with the isotropy of
the microwave background within the stringent limit of one part in
$10^5$. Indeed, the energy-momentum tensor, $T_{\mu \nu}$, associated to
this solution has the  spatial components $T_{mn}$  proportional to
${\delta}_{mn}$.

The energy density, $\rho_w$, stored in this configuration of scalar fields
is:

\eqn{rho_w}{ \rho_w = {{\nabla}_i{\varphi}_j {\nabla}^i{\varphi}_j\over
2a^2}={3  w^2 f^2\over 2 L^2a^2}}
This energy density redshifts like the spatial curvature contribution to
the Friedmann equation, however it does not alter the relation between
angular and radial distances from their flat space behaviour.

The same dependence of the energy density on the scale factor can be
obtained for an ideal fluid with the following equation of state :

\eqn{-1/3rho}{P = -{1\over 3}\rho}

It is amusing to note that this equation lies at the
boundary between equations of state that lead to an accelerating
universe and the ones that lead to the deccelerating case. However, in our
script as will become clear, the energy stored in the "wound" scalar
fields has never been the dominant component of the energy density of the
universe. Therefore that component of the energy density does not affect
the leading characteristics of the expansion of the universe.

  Next, we will discuss what the requirements on the various parameters
are, in order for this energy density to survive to the
present day and contribute at best one percent of the total energy
density.

Before that, we want to briefly address the reader that may 
 wonder whether these massless scalars
generate long range forces  that would experimentally rule them out.
 A simple answer to such concern, is to require that
these scalars not couple directly to the low energy degrees of freedom, except 
through gravity. In addition, we should also stress that even in the absence of 
such restriction, these scalars couple derivatively with a "decay constant" that is 
close to the Planck scale, as will be seen next, and are therefore not
expected to affect static forces between bulk matter at macroscopic distances.

\section{Constraints on Parameter Space}

\subsection{The requirements on the parameters related to the torus and
scalars}

There are various bounds on the spatial size of the universe when the
universe is assumed to have the topology of a torus. The most stringent
bound, to our knowledge, comes from the absence of "circles in the sky"
\cite{Cornish:2003db} and requires the physical size of the universe to be
no smaller than about six times the size of our present horizon.

\eqn{sizebound}{La_0\geq 24  \mbox{ Gpc} \sim 5 \times 10^{61}l_P}
where $l_P= ( \hbar G / c^3 )^{1/2} $ is the Planck length.

We can now proceed to estimate the energy density, ${\rho}_w$ stored today
in the "wound" scalar fields as a function of the parameter f, "the decay
constant" of the ${\theta}_i$.

\eqn{rhow}{{\rho}_w={3w^2f^2\over 2L^2a^2}\leq {3w^2f^2\over 5\times
10^{123}{l_P}^2}}
which gives a contribution to the density of the universe today
\eqn{omegachi}{ \frac {\rho_{w}}{\rho_{{cr}}} = \left(\frac { f  \,\, 
w} { 2 \times 10^{19} \mbox{GeV}}\right)^2}
A one percent contribution to the present energy density corresponds to
$f\sim 2 \times  10^{18}$ GeV.  Though, this decay constant is
close to the Planck scale, it doesn't invalidate the use of classical
gravity to describe the evolution of the universe. Indeed the 
contribution to the Ricci
scalar from this form of energy is given by:

\eqn{RR}{ R  \, l_{P}^2 \sim 24 \p  \left( \frac{l_P}{ L a(t)} \right 
) ^2 \left( \frac{w \,f}{1.2 \times 10^{19} \mbox{GeV}} \right)^2 }
Therefore as long as the physical size of the universe exceeds $10 
l_P$, the scalar curvature is safely subplanckian.

It is worth reminding the reader that we imposed earlier a
restriction on parameter space by choosing the same
"decay constant",
$f$ , for the ${\theta}_i$ to insure the isotropy of the microwave background.

There are also various other requirements on the "decay constant",  $f$, the
size of the torus, $L$, as well as on the parameters related to 
inflation: the energy
density during inflation $\Lambda_I$, the number of e-foldings, $N_e$, ...
which will be discussed next.

\subsection{The requirement on the parameters coming from inflation}

In the presence of this additional form of energy, the onset of 
inflation depends on the energy density stored in the scalar
fields to become smaller than the energy density, $\Lambda_{I}$,
which sets the energy density scale of inflation.

\eqn{inflation}{\rho_w(t_{\mbox{beginning inflation}}) << \Lambda_{I}}
This constraint can be expressed as a function of the $N_e$ and the 
reheating temperature, $T_R$,

\eqn{inflationreh}{ \rho_{w}(t_0)<< \Lambda_{I}  e^{- 2 N_e}\left(\frac{T_0}{T_R} 
\right)^2 }

Another important constraint comes from requiring that the fluctuations
generated during inflation, in the energy density of the scalar field
(in the presence of their non-trivial topological background) not
overwhelm the scale invariant fluctuations produced by the inflaton.
The perturbations in the wound  fields ( $\delta \theta_i $ ) 
together with the perturbations on
the inflaton field  ( $\delta  \phi $)  will generate a perturbation 
on the stress energy
density. In turn, these perturbations will generate a perturbation in
the metric ($ \Psi$). To relate this latter variable to first two, we 
should construct the linear combination
of these quantities that is conserved outside the horizon. It is given by
\cite{Bardeen:1980kt,Dodelson:2003}

\eqn{bardeen}{ \zeta= \frac{- i k_i \delta T^0_i H }{ k^2 ( \rho  + P 
)} - \Psi }
One can separately analyze the contributions to $\delta T^0_i $ from 
the two types
of fields.

\eqn{emtwoundfields}{\delta {T_{w}^{\,\,\,\,0}}_i= - \frac{   w f^2} 
{L } \frac{ d \delta \theta_i}{dt}}

\eqn{emtinflaton}{ \delta {T_{\phi}^{ \,\,\,\,0}}_i=- i k_i \frac{d 
\phi}{dt} \delta \phi }
Since  $|\frac{d\delta \theta_i}{dt}|= \frac{H}{\sqrt{2}} |\delta 
\theta_i|$ at horizon crossing, the fluctuations around
the wound fields, if dominant, will generate a non scale invariant 
power spectrum. Both
the inflaton field and the wound fields are scalars and will have the 
same quantum perturbations.

\eqn{quantumpert}{<\delta \phi({\bf k},t) \delta \phi(-{\bf k},t)>= 
f^2 <\delta \theta_i({\bf k},t) \delta \theta_i(-{\bf k},t)>}
Hence the condition to avoid violations of scale invariance is that:

\eqn{fluctuationcondition}{\rho_{w} << (P_{\phi} + \rho _ {\phi}) = 
\left(\frac{d \phi}{dt}\right)^2}
This condition, in turn, ensures that the contribution to $\Psi$ from 
the wound fields is
negligible at sub-horizon and crossing horizon scales.

\section{Experimental information}

In this section we explore the observational constraints on the 
various inflationary parameters discussed in the previous section. In addition, we also discuss the extent to which the $1/a^2$ contribution to the energy density could be present today.
Recent WMAP data  \cite{Peiris:2003ff} yields values
\eqn{tensortoscalarratio}{r(k_0=0.002 \, \mbox{Mpc} ^{-1})<0.90}
\eqn{scalarA}{A(k_0=0.002 \, \mbox{Mpc} ^{-1})=0.75^{+0.08}_{-0.09}}
where $r$ is the relative amplitude of the tensor to scalar modes
\eqn{tsr}{r\equiv\frac{\Delta_h^2(k_0)}{\Delta_R^2(k_0)}}
and $A(k_0)$ and $\Delta_R^2(k_0)$ are related through
\eqn{scalarspectrum}{\Delta_R^2(k_0)\simeq2.95\times10^{-9}A(k_0)}
Standard slow roll analysis to the first order gives:
\eqn{scalarslowroll}{\Delta_R^2(k_0)=\frac{V/M_P^4}{24\pi^2\epsilon_V}}
\eqn{tsrslowroll}{r=16\epsilon_V}
where the slow roll parameter $\epsilon_V$ is defined by:
\eqn{epsilon}{\epsilon_V\equiv\frac{M_P^2}{2}\left(\frac{V'}{V}\right)^2}
\eqn{Planckmass}{M_P\equiv(8\pi G)^{-1/2}=2.4\times10^{18} \, \mbox{GeV}}
During inflation
\eqn{Hinfl}{H^2\simeq\frac{V}{3M_P^2}}
\eqn{Vprime}{3H\dot{\phi}\simeq-V'}
so the WMAP data corresponds to
\eqn{phidot}{\dot{\phi}\simeq2\times10^{-34} \frac{V} {\mbox{GeV}^{-2}}}
\eqn{Vconstraint}{V<(3.1\times10^{16} \, \mbox{GeV})^4}
In the previous section we have obtained the requirement 
(\ref{fluctuationcondition}):
\eqn{rho_wconstraint}{\rho_w<\dot{\phi}^2}
Assuming that the modes of the size of present horizon left the
horizon during inflation $N_e$ e-foldings before the end of the
inflation, which in our scenario is the beginning of inflation, and 
also assuming maximally efficient reheating, we obtain:
\eqn{f}{f<1.2\times10^{-47}e^{-N_e}V^{3/4}La_0}
(To make this estimate, we use $\rho_w< 
\frac{1}{10}\dot{\phi}^2$.) 
In particular, taking ${La_0\sim24 \, \mbox{Gpc}}$, ${V\sim(3\times10^{16} \,\mbox{GeV})^4}$ and $N_e=60$ 
we obtain the limit
\eqn{fconstraint}{f<10^{19} \, \mbox{GeV}}

There are additional constraints from the observational data which can 
potentially limit the contribution of $1/a^2$ term to the present energy 
density. In particular, recent high-redshift supernova measurements 
\cite{Riess:2004nr} seem to allow flat Universe with e.g. 
$\Omega_{matter}\simeq0.27$, $\Omega_{\Lambda}\simeq0.53$ and 
$\Omega_w\simeq0.20$, within the $95\%$ confidence level ($\Omega_w$ 
being the contribution of $1/a^2$ term to the present energy 
density).
A considerable $\Omega_w$ would also have an effect on the age of the 
Universe. In particular, for a flat Universe with 
$\Omega_{matter}=0.27\pm0.04$ \cite{Peiris:2003ff}, $\Omega_w=0.2$, 
$\Omega_{radiation}=10^{-4}$ and 
$\Omega_{\Lambda}=1-\Omega_{matter}-\Omega_{radiation}-\Omega_w$, 
using the HST Key Project \cite{Freedman:2000cf} value of 
$H_0=72\pm3\pm7 \, \mbox{km} \,\mbox{s}^{-1}\,\mbox{Mpc}^{-1}$ one obtains the age of the Universe 
to be $13\pm2 \,\mbox{Gyr}$ which is in line with the present observational data 
\cite{Cowan:1998nv,Hansen:2002ij,NeillReid:1997,Bennett:2003bz}.  
It is probable, that as suggested to us by E. Komatsu, the growth rate of density 
fluctuations and the mass function of collapsed objects will be useful to put more stringent constraints on $\Omega_w$.
\vfill\eject

\includegraphics[width=7in]{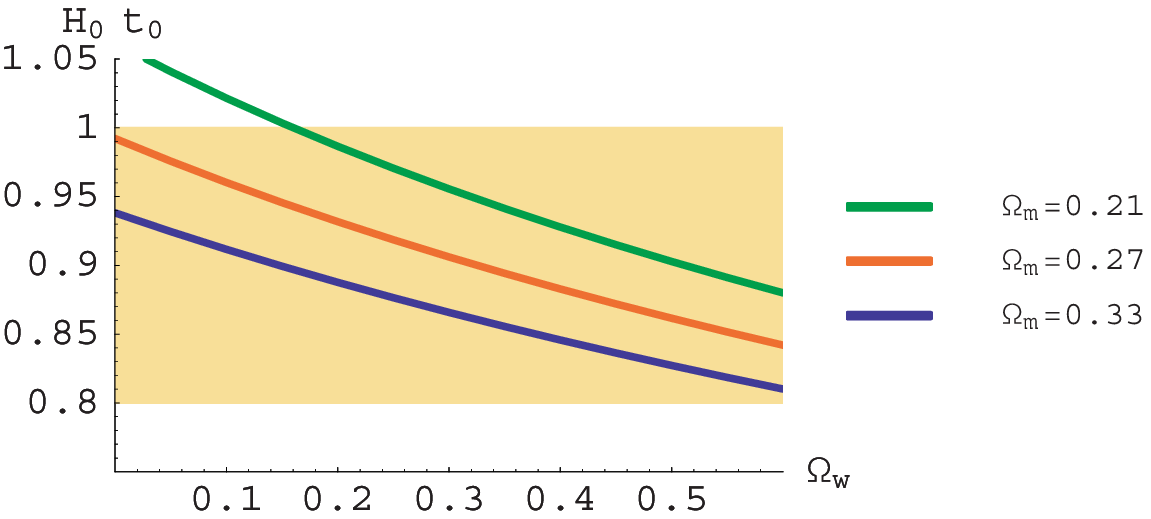}

\centerline{The effect of $\Omega_w=1-\Omega_m-\Omega_\Lambda$ on the
age of the Universe. }
\centerline{The light yellow region corresponds to the globular cluster age $12.7\pm0.7 \,
\mbox{Gyr}$ \cite{Hansen:2002ij}.}

  \section{Conclusions}

The experimental data seem to leave open the possibility of a 
contribution to the present energy that redshifts like $1/a^2$.
We present in this paper an amusing and extremely constrained setup 
that cannot yet be excluded experimentally and does yield such a
contribution to today's energy density.

There are, in addition, other known sources 
\cite{Spergel:1996ai,Copeland:2003bj,Kamionkowski:1996tj} for
a component of the energy density that behaves like  $1/a^2$. It is
unclear, however, whether these sources would contribute in
such manner today \cite{McGraw:1997nx}. Our hope, is that the experiments will put firmer 
bounds on the existence of a $1/a^2$ contribution to the present
energy density, irrespective of what sources such contribution.

\acknowledgments We would like to thank Raphael Flauger, Eiichiro 
Komatsu, Chethan Krishnan and Hyukjae Park for useful discussions.
This material is based upon work supported by the National Science Foundation under Grant No. 0071512.


\end{document}